\documentstyle[prl,aps,epsfig,multicol]{revtex}

\newcommand{\sst}{\scriptscriptstyle}

\def\si{^1 \hskip -0.03in S _0}
\def\siii{^3 \hskip -0.025in S _1}
\def\pii{^1 \hskip -0.03in P _1}
\def\piii{^3 \hskip -0.03in P _0}
\def\piiii{^3 \hskip -0.025in P _1}

\begin{document}
\title{The $\Lambda N \to NN$ weak interaction in effective field theory}
\author{Assumpta Parre\~no$^a$, Cornelius Bennhold$^b$, Barry R. Holstein$^{c,d}$}
\address{$^a$Departament ECM, 
Facultat de F\'{\i}sica, Universitat de Barcelona, E-08028, Barcelona, Spain
}
\address{$^b$Center of Nuclear Studies, Department of Physics, 
The George Washington University,
Washington DC, 20052, USA}
\address{$^c$Department of Physics-LGRT, University of Massachusetts,
Amherst, MA 01003, USA}
\address{$^d$Thomas Jefferson National Accelerator Facility, Theory Group, 
Newport News, VA 23606 }
\date{\today}

\draft
\maketitle

\begin{abstract}

The nonleptonic weak $|\Delta S|=1$ $\Lambda N$ interaction, responsible
for the dominant nonmesonic decay of all but the lightest hypernuclei, is
studied in the framework of an effective field theory. The long-range
physics is described through tree-level exchange of the SU(3) Goldstone
bosons while the short-range potential is parametrized in terms of
lowest-order contact terms.
We obtain reasonable fits to available weak hypernuclear decay
rates and quote the values for the parity-violating asymmetry as predicted by the
present effective field theory.
\end{abstract}

\pacs{PACS numbers: 21.80.+a, 13.30.Eg, 13.75.Ev}

\begin{multicols}{2}


For the past 50 years, $\Lambda$-hypernuclei, systems of one or more
$\Lambda$ hyperons bound to a core nucleus, have been used to extend our
knowledge of both the strong and the weak baryon-baryon
interaction from the $NN$ case into the SU(3) sector.
The nonmesonic hypernuclear weak
decay, facilitated via the $|\Delta S|=1$ four-fermion interaction,
thus complements the weak $\Delta S=0$ $NN$ case, which allows the study of
only the parity-violating amplitudes.

In analogy to $NN$ phenomenology, the nonmesonic hypernuclear decay
has traditionally been modeled using a meson-exchange
approach\cite{mes}. The long-range
part of the interaction is naturally explained by one-pion exchange
which could approximately reproduce the total (one-nucleon induced)
nonmesonic decay rate, $\Lambda N \to NN$,
but not the partial rates, the proton-induced $\Lambda p \to np$ rate
$\Gamma_p$ and the neutron-induced
$\Lambda n \to nn$ rate $\Gamma_n$\cite{AG02}.
Due to the $\Lambda N$ mass difference,
 the process $\Lambda N \to NN$ produces nucleons
with momenta around $\approx$ 420 MeV, suggesting that the short-range
part of the interaction cannot be neglected.  These contributions have been described either
through the exchange of vector mesons\cite{PRB97,DFHT96},
whose production thresholds are too high for the free $\Lambda$ decay,
 or through direct quark exchange\cite{ISO00}.

In contrast to previous theoretical studies, we present an exploratory study
in order to determine the possible efficacy of the use of Effective Field 
Theory (EFT) methods in
hypernuclear decay. Studies in
this direction have already begun in Ref.~\cite{Jung} where a
Fermi (V-A) interaction was added to the OPE mechanism to describe
the weak $\Lambda N \to NN$ transition. The present approach is motivated 
by the remarkable success of EFT techniques based on chiral 
expansions in the (non-strange)
SU(2) sector\cite{BBHPS,UvK99,ZPHR01,Ulf}, which suggests its
extension to the SU(3) realm, even though stability of the chiral
expansion is less clear for the SU(3) sector, due to the
significant degree of SU(3) symmetry breaking.
A well-known example of the problems facing SU(3) chiral
perturbation theory has been the
prediction\cite{jenkins92,springer99,borasoy03,BH99:nonlep} of the
four parity-conserving (PC) amplitudes in the weak nonleptonic decays
of octet baryons, $Y \to N \pi$, with $Y=\Lambda, \Sigma$ or
$\Xi$. In particular, Refs. \cite{BH99:nonlep} and \cite{BH99:rad}
studied the contributions from negative-parity intermediate states
and demonstrated their potential to resolve this longstanding
issue.  For our purpose, these higher-order effects are beyond the
simple lowest-order analysis considered here.

The EFT approach is based on the existence of well separated
scales in the physical process under study. Formally, the
high-momentum (short-distance) modes in the four-baryon
interaction Lagrangian are replaced by contact operators,
compatible with the underlying physical symmetries, of increasing
dimension. Built in such way, the Lagrangian will contain an
infinite number of terms, and a consistent power-counting scheme
is needed in order to truncate such an expansion to a given order.
The coefficients appearing in the Lagrangian are then fitted to
reproduce the available data in the low-energy regime. Whether an
EFT will succeed to describe a particular process or not is
directly related to the success in obtaining a controlled,
systematic expansion in terms of a small parameter.
In contrast to the $NN$ case, however, the $\Lambda N \to NN$
transition corresponds to an energy release $\approx 177$ MeV
($|{\bf p}| \approx 417$ MeV) at threshold. It is therefore not at
all clear if low-energy expansions can be successfully carried
out. In light of this energy release at threshold, it is
reasonable to include the pion ($m_\pi \approx 138$ MeV) and the
kaon ($m_K \approx 494$ MeV) as dynamical fields. Working within
SU(3) also supports treating the pion and kaon on equal footing.
The last member of the SU(3) Goldstone-boson octet, the $\eta$, is
usually not included, since the strong $\eta NN$ coupling is an
order of magnitude smaller than the strong $\pi NN$ and $K \Lambda
N$ couplings\cite{etacoup}. Thus, following a power-counting
scheme based on the engineering dimensions of the operators, at
leading order (order unity in the external momentum, $p^0$\, ) the
present study of the weak $\Lambda N \to NN$ transition includes
the contribution of the long-ranged pion and kaon exchanges, and a
short-range contribution given by leading nonderivative contact
terms (leading order parity-conserving, LO PC, terms). Our study
also includes next-to-leading order (NLO) terms in the momentum
expansion (or equivalently, leading order parity-violating pieces,
LO PV).
Not considered here are contributions from intermediate-range
2$\pi$-exchange. Such a component is two orders higher in the
chiral expansion than the corresponding single pion-exchange
piece.  Also, previous studies\cite{JOP01} found such
contributions to be small due to significant cancellations between
the correlated and uncorrelated 2$\pi$-exchanges.

The weak and strong Lagrangians for pion and kaon
exchanges are the same as in Ref. \cite{PR01}, and
we use experimental values for the couplings at the strong
($g_{\rm{\sst NN} \pi}=13.16$)
and weak ($\Lambda N \pi = 1.05$ for the parity-violating
amplitude and $-7.15$ for the parity-conserving one, in units of
$G_F {m_\pi}^2 = 2.21 \times 10^{-7}$) baryon-baryon-pion vertices.
Like the $NN\pi$ coupling, the $\Lambda N
K$ and $\Sigma N K$ coupling constants represent a fundamental input
into our calculation. Unlike the $NN\pi$ coupling, however, their
values are considerably
less well known, with $g_{\rm{\sst \Lambda N K}} = 13 -
17 $ and $g_{\rm{\sst \Sigma N K}} = 3-6 $.  Here, we choose the
values given by the Nijmegen Soft Core NSC97f interaction 
model\cite{nij99},$g_{\rm{\sst
\Lambda N K}} = -17.66$ and $g_{\rm{\sst \Sigma N K}} = -5.38$.
Of course, the weak nucleon-nucleon-kaon
coupling constants are not accessible
experimentally, so we obtain numerical values by making
use of SU(3) and chiral algebra considerations\cite{PRB97,DFHT96}.
In addition, those potentials will be regularized by using monopole
form factors at each vertex\cite{ff}.
We note that all the strong model-dependent
ingredients used in the present calculation (such as cut-off parameters
or strong couplings)
have been taken from the Nijmegen Soft-Core model f\cite{nij99}.


If no model is assumed, the low energy $\Lambda N \to N N$ process can be
parametrized through the 6 partial waves listed in Table I\cite{comment1}.
The $\si \to$ $\si$ and $\siii \to$ $\siii$ transitions can only be produced
by the ${\hat 1} \, \cdot \, \delta^3({\vec r} \,)$ and
${{\vec \sigma_1} \cdot{\vec \sigma_2}} \, \cdot \, \delta^3({\vec r} \,)$
operators,
where $\delta^3({\vec r} \,)$ represents the contact interaction.
The $\si \to$ $\piii$ and $\siii \to$ $\pii$ transitions proceed through
the combination of the spin-nonconserving operators:
$({\vec \sigma_1} - {\vec \sigma_2})
\cdot \{ \vec{p}_1 - \vec{p}_2 \, , \, \delta^3({\vec r} \,) \, \}$, \,
$({\vec \sigma_1} - {\vec \sigma_2}) \cdot
[ \, \vec{p}_1 - \vec{p}_2\, , \, \delta^3({\vec r} \,) \, ]$,
$\,\, {\rm i} \, ({\vec \sigma_1} \times {\vec \sigma_2}) \cdot \{
\vec{p}_1 - \vec{p}_2 \, , \,
\delta^3({\vec r} \,) \, \}$ and ${\rm i} \, ({\vec \sigma_1} \times
{\vec \sigma_2}) \cdot
[ \, \vec{p}_1 - \vec{p}_2\, , \, \delta^3({\vec r} \,) \,]$,
where $\vec{p}_i$ is the derivative operator acting on the
"i{\it th}" particle
\cite{comment2}.
The \mbox{$\siii \to$ $\piiii$} transition is allowed by the combination of
the spin-conserving operators
$({\vec \sigma_1} + {\vec \sigma_2})
\cdot \{ \vec{p}_1 - \vec{p}_2 \, , \, \delta^3({\vec r} \,) \, \}$ and
$({\vec \sigma_1} + {\vec \sigma_2}) \cdot [ \, \vec{p}_1 - \vec{p}_2
\, , \, \delta^3({\vec r} \,) \,]$,
while only two-derivative operators can produce the last
(tensor) transition.

Using power counting we can discard operators of order $q^2/{M_N}^2$,
where the momentum transferred is defined by ${\vec q} = {\vec p}_\Lambda -
{\vec p}_{N_1}= {\vec p}_{N_2} - {\vec p}_N$.
The remaining (lowest order) operators lead to the following four-fermion
interaction in ${\vec r}$-space
(in units of $G_F = 1.166 \times 10^{-11}$ MeV$^{-2}$):
\begin{eqnarray}
V_{4P} ({\vec r}) &=&
\left\{C_S^0 + C_S^1 \; {\vec \sigma}_1 \cdot{\vec
\sigma}_2\right. \nonumber\\
&+& \left.\displaystyle\frac{2 r}{\delta^2}
\left[ C_P^0 \; \displaystyle\frac{{\vec \sigma}_1 \cdot{\hat r}}{2 {\overline M}}
+ C_P^1 \; \displaystyle\frac{{\vec \sigma}_2 \cdot{\hat r}}{2 M}
+ C_P^2 \; \displaystyle\frac{({\vec \sigma}_1 \times {\vec \sigma}_2)
\cdot {\hat r}}{2 {\tilde{M}}} \right]\right\} \nonumber \\
&\times& \displaystyle\frac{e^{- \frac{r^2}{\delta^2}}}{\delta^3 \pi^{3/2}}
\; \times \left[
C_{IS} \, {\hat 1} + C_{IV} \,
{\vec \tau_1}\cdot {\vec \tau}_2
\right] \, {\rm ,}
\label{eq:4ppotr}
\end{eqnarray}
where the last factor represents the
$\Delta I=1/2$
isospin part of the 4P interaction. Note that
the delta functions have been smeared by using a Gaussian form with
a typical vector-meson ($\rho$) range,
$\delta \sim \sqrt{2} m_\rho^{-1} \approx 0.36$ fm.
Here, $\tilde{M}= (3 M+M_\Lambda)/4$ is a weighted average
of $N,\Lambda$ masses while 
$C^j_S$ and $C^j_P$ are
the j$th$ Low Energy Coefficients (LEC) at 
0$th$ and first order respectively. Although
the form of the
contact terms is model-independent, the size of these LEC's depends
upon how the theory is formulated,
and they are expected to be of order of the other couplings in the problem.
These couplings provide a very simple representation of the short
distance contributions to the process at hand.  In a complete model,
they would be represented by specific dynamical contributions, such as
$\rho,\omega, etc.$-exchange.  However, we eschew the temptation to be
more specific---in fact this generality is one of the strengths of our
approach. We evaluate the coefficients purely phenomenologically and
leave theoretical interpretation of the pieces to future
investigations.  Of course, the specific size of such coefficients depends upon
the chiral order to which we are working.  However, if the expansion
is convergent, then the values of these effective couplings should be
relatively stable as NLO or higher effects are included.


It is well known that the high momentum transferred in the $\Lambda N \to NN$
reaction makes this process sensitive to the short range
physics which is characterized by our contact coefficients.
Moreover, since the $|\Delta S|=1$ reaction takes place in a finite nucleus,
the extraction of reliable information of the elementary weak two-body
interaction requires a careful investigation of the many-body nuclear
effects present in the hypernucleus.
In the present calculation, we use a shell-model for the initial hypernucleus,
where the single-particle $\Lambda$ and N orbits
are taken to be solutions of harmonic oscillator mean field potentials
with parameters ($b_\Lambda$ and $b_{\rm N}$)
adjusted to experimental separation energies
and charge form factor (respectively) of the hypernucleus under study.
For $^{12}_\Lambda$C and $^{11}_\Lambda$B
$b_{\rm N} = 1.64$ fm and $b_\Lambda = 1.87$ fm, while for
$^5_\Lambda$He they take the values $b_{\rm N} = 1.4$ fm and
$b_\Lambda = 1.85$ fm. The strong YN interaction at short distances,
absent in mean-field models, is accounted for by replacing the
mean-field two-particle $\Lambda N$ wave function by
a correlated\cite{sitges} one
obtained from a microscopic finite-nucleus $G$-matrix
calculation \cite{Ha93} using the soft-core and hard-core
Nijmegen models \cite{NR77}.
The $NN$ wave function is obtained from the
Lippmann-Schwinger ($T$-matrix) equation with the input of the
Nijmegen Soft Core potential model; details of the calculation
can be found in the
Appendix of Ref. \cite{PR01} which presented a detailed study of different approaches to
Final State Interactions (FSI) in the decay process.


We begin the discussion of our results with a remark on the data.
One might wonder if there can be only three independent data points
in the nonmesonic
decay: the proton-induced and neutron-induced rates $\Gamma_p$ and $\Gamma_n$,
and the asymmetry $\cal {A}$ (associated with the proton-induced decay),
relating observables from one hypernucleus to another through hypernuclear
structure coefficients. While one may
indeed expect measurements from different p-shell hypernuclei, say, A=12 and 16,
to provide the same constraint, the situation is different when including data
from s-shell hypernuclei like A=5. For the latter, the initial $\Lambda N$
pair can only be in a relative s-state, while for the former, relative p-states
are allowed as well.  We therefore
include data from the A=5,11 and 12 hypernuclei in our fits.

Note that we do not attempt to perform a final, quantitative
fit to all hypernuclear data. Rather, we are exploring whether an
EFT can be used to reproduce various reasonable subsets of the
hypernuclear decay data in order to verify the validity of such an
expansion. We note that the presently large experimental error bars in hypernuclear decay
observables puts strong limitations in any EFT approach to the decay process. 
In this sense, the values for the parameters we are
presenting have to be taken with caution. A quantitatively more
rigorous understanding awaits better data as well as theory which
includes higher-order effects. Only the more recent measurements
from the last 12 years were used, excluding, however, those
$\Gamma_n$/$\Gamma_p$ (hereafter $= n/p$) ratio data whose error
bars were larger than 100$\%$.   A more detailed discussion of the
minimization calculation will be presented
elsewhere\cite{PBHlong,minuit}.

We also have taken the data at face value and have not applied any
corrections due to, {\it e.g.}, the two-nucleon induced mechanism,
which has been estimated to amount up to $25\%$ of the total decay
rate for p-shell hypernuclei\cite{2N,GPR04}. Ideally, exclusive
experiments would separate this mechanism from the measured
total decay rate, permitting a fit to observables that are not
contaminated with multi-nucleon effects.  At the present time,
given the sizable error bars of the data, this omission would not
change our conclusions.

No parameters were fitted for the results with only
$\pi$ and $K$ exchange, shown in Table III.  As has been known for a long time, 
$\pi$ exchange
alone reasonably well describes the observed total rates, while dramatically
underestimating the $n/p$ ratio.
The tensor PC channel dominates the proton-induced
rate while it is absent
in the $L=0$ neutron-induced one.
Incorporation of kaon exchange yields a destructive interference between both
mechanisms (OPE and OKE) in the PC amplitudes,
while the interference is constructive in their PV
counterparts.
As a consequence,
$n/p$ is enhanced by about a factor of five, within reach of the lower
bounds of the experimental
 measurements, while the total rate underpredicts the observed value
by about a factor of two. This $\pi-K$ interference also leads to
values for the asymmetry that are close to experiment for the p-shell
hypernuclei, but far off for A=5\footnote{Note that for $^5_\Lambda$He we
quote the value of the intrinsic $\Lambda$ asymmetry parameter, $a_\Lambda$, 
which is experimentally accessible, while for p-shell hypernuclei the accessible
quantity is $\cal{A}$, the difference between the number of protons coming parallel and 
antiparallel with respect to the polarization axis. This quantity can be 
related to $\alpha_\Lambda$ through the relation ${\cal A} = p_\Lambda a_\Lambda$,
where $p_\Lambda$ is the $\Lambda$ polarization, to be extracted
from theoretical models.}. 
Since the contributions of both
$\eta$-exchange and two-pion exchange are negligible, these discrepancies
illustrate the need for short-range physics.

Allowing contact terms of order unity (leading-order PC operators)
to contribute leads to four free parameters, $C_S^0$, $C_S^1$,
$C_{IS}$ and $C_{IV}$. Data on the total and partial decay rates
for all three hypernuclear systems are included in the fit, but no
asymmetry measurements. The inclusion of the contact terms roughly
doubles the values for the total decay rates, thus restoring
agreement with experiment. The impact on the $n/p$ ratio is
noteworthy: the value for $^5_\Lambda {\rm He}$ increases by
10$\%$ while the $n/p$ ratios for $^{11}_\Lambda {\rm B}$ and
$^{12}_\Lambda {\rm C}$ almost double. This is an example of the
differing impact certain operators can have for s- and p-shell
hypernuclei. The effect on the asymmetry is opposite, almost no
change for A=11 and 12, but a 30$\%$ change for A=5. The
magnitudes of the parameters, $C_S^0$, $C_S^1$, 
$C_{IV}$, listed in Table II, are each around their natural size
of unity, while $C_{IS}$ is a factor of three or so larger. Note the
substantial error bars on all the parameters, reflecting the
uncertainties in the measurements.

Three new parameters are admitted when we allow the leading-order PV terms
(of order $q/M_N$) to
contribute with the coefficients $C_P^0$, $C_P^1$, and $C_P^2$.
As shown in Table II, the parameters for the PV contact terms are
larger than the ones for the PC terms, and in fact, compatible
with zero. Including the three new parameters does not
substantially alter the previously fitted ones, thus supporting
the validity of our expansion. Regarding their impact on the
observables, the PV contact terms barely modify the total and
partial rates but significantly affect the asymmetry, as one would
expect for an observable defined by the interference between PV
and PC amplitudes. The calculated 
asymmetry changes sign for all
three hypernuclei, moving the $^5_\Lambda {\rm He}$ value within
the measured range at the expense of the one for $^{11}_\Lambda
{\rm B}$. This shift occurs without any asymmetry data
constraining the fit. In order to further understand this
behavior, we have performed a number of fits including the
asymmetry data points of either $^5_\Lambda {\rm He}$ or
$^{11}_\Lambda {\rm B}$ or both. Tables II and III display the
result of one of those fits. Inclusion of the $^5_\Lambda {\rm
He}$ asymmetry helps in constraining the values of two of the LO
PV parameters. We find that the two present experimental values
for A=5 and A=11 cannot be fitted simultaneously with this set of
contact terms. Future experiments will have to settle this issue.

We have also performed fits allowing a contribution from an isospin
$\Delta I = 3/2$ transition operator.
The resulting fit, with $\hat{\chi}^2 \sim 1.4$, shifts strength from the
isoscalar contribution to the new $\Delta I = 3/2$ one, leaving the
other parameters unchanged. However,
as shown in Table II,
we can clearly get an good fit to all observables without such
transitions while obtaining couplings of reasonable size.
In addition,
we have checked that
our conclusions are independent of the strong interaction
model used to describe FSI in the
transition. Employing $NN$ wave functions that are obtained with
either the NSC97f or the NSC97a model in the fit leaves the
observables almost unchanged, with the exception of the asymmetry parameter,
which can change up to 50\%. The obtained couplings can easily absorb the
changes but remain compatible within their error bars. Similarly,
we performed a study of the sensitivity of the calculated
observables to the smearing function in Eq.~(1). For
values of $\delta$ going from $0.3$ to $0.4$ fm,
the results are remarkably insensitive,
except again in the case of the asymmetry.


In conclusion, we have studied the nonmesonic weak decay using an
Effective Field Theory framework for the weak interaction.  The
long-range components were described with pion and kaon exchange,
while the short-range part is parametrized in leading-order PV and
PC contact terms.  We find coefficients of natural size with
significant error bars, reflecting the level of experimental
uncertainty.  The largest contact term corresponds to an
isoscalar, spin-independent central operator.  There is no
indication of any contact terms violating the $\Delta I$ = 1/2
rule. In this study we have not speculated as to the dynamical
origin of these contact contributions.  Rather our aim was to
ascertain their basic magnitude and to establish the validity of
the EFT framework for the weak decay. The next generation of data
from recent high-precision weak decay experiments currently under
analysis holds the promise to provide much improved constraints
for studies of this nature.

The authors are indebted to Daniel R. Phillips, Angels Ramos,
Martin J. Savage and Roxanne P. Springer
for several helpful and encouraging discussions.
Work supported by EURIDICE HPRN--CT--2002--00311,
DGICYT BFM2002--01868, Generalitat de Catalunya
SGR2001--64,
U.S. DOE grant no. DE-FG02-95ER-40907,
and U.S. NSF PHY-02-44801.

\vspace*{1cm}
\begin{table}[hbt]
\caption{$\Lambda N \to NN$ partial waves.}
\begin{tabular}{lccc}
{\rm partial \,\, wave} & {\rm operator} & {\rm order} & I \\
\hline
$^1S_0 \to ^1S_0$ & ${\hat 1}, \, {\vec \sigma_1} {\vec \sigma_2}$
& $1$ & $1$ \\
$^1S_0 \to ^3P_0$ & $({\vec \sigma_1} - {\vec \sigma_2}) {\vec q} {\rm ,}
\,\, ({\vec \sigma_1} \times {\vec \sigma_2}) {\vec q}$ & $q/M_N$ & $1$ \\
$^3S_1 \to ^3S_1$ & ${\hat 1}, {\vec \sigma_1} {\vec \sigma_2}$ & $1$ & 0 \\
$^3S_1 \to ^1P_1$ & $({\vec \sigma_1} - {\vec \sigma_2}) {\vec q} {\rm ,}
\,\, ({\vec \sigma_1} \times {\vec \sigma_2}) {\vec q} $ & $q/M_N$ & $0$ \\
$^3S_1 \to ^3P_1$ & $({\vec \sigma_1} + {\vec \sigma_2}) {\vec q}$
& $q/M_N$ & $1$ \\
$^3S_1 \to ^3D_1$ & $({\vec \sigma_1} \times {\vec q}) ({\vec \sigma_2}
\times {\vec q})$ & $(q/{M_N})^2$ & $0$ \\
\end{tabular}
\end{table}

\begin{table}[hbt]
\caption{LEC coefficients corresponding to the LO
calculation.
The values in parentheses have been obtained including
${\alpha_\Lambda} (^5_\Lambda {\rm He})$ in the fit.}
\begin{tabular}{l|c|c}
 & $+$ LO (PC) & $+$ LO (PC+PV) \\
\hline
$C_S^0$ & $-1.54 \pm 0.39$  & $-1.31 \pm 0.41
\, (-1.04 \pm 0.33)$ \\
$C_S^1$ & $-0.87 \pm 0.24$ & $-0.70 \pm 0.35
\, (-0.57 \pm 0.27)$ \\
$C_P^0$ & $---$ & $-5.82 \pm 5.31
\, (-4.49 \pm 1.57)$ \\
$C_P^1$ & $---$  & $2.47 \pm 3.13
\, (1.84 \pm 1.93)$ \\
$C_P^2$ & $---$ & $-5.68 \pm 3.13
\, (-4.47 \pm 2.31)$ \\
$C_{IS}$ & $5.01 \pm 1.26$ & $4.68 \pm 0.67
\, (5.97 \pm 0.86)$ \\
$C_{IV}$ & $1.45 \pm 0.38$ & $1.22 \pm 0.20
\, (1.56 \pm 0.26)$ \\
\end{tabular}
\end{table}

\end{multicols}
\begin{table}[hbt]
\caption{Results obtained for the weak decay observables,
when a fit to the $\Gamma$ and $n/p$ for
$^5_\Lambda {\rm He}$, $^{11}_\Lambda {\rm B}$ and
$^{12}_\Lambda {\rm C}$ is performed.
The values in parentheses have been obtained including
${\alpha_\Lambda} (^5_\Lambda {\rm He})$ in the fit.}
\begin{tabular}{l|c|c|c|c|c}
 & $\pi$ & $+ K$ & $+$ LO & $+$ LO & EXP: \\
 &  &  & PC & PC + PV & \\
\hline
$\Gamma (^5_\Lambda {\rm He})$ & $0.42$ & $0.23$
& $0.43$  & $0.44$ ($0.44$)
 & $0.41 \pm 0.14$\cite{Szy91}, \, $0.50 \pm 0.07$\cite{No95}  \\
$n/p (^5_\Lambda {\rm He})$ & $0.09$ & $0.50$
& $0.56$ & $0.55$ ($0.55$)
& $0.93\pm 0.55$\cite{Szy91}, \, $0.50 \pm 0.10$\cite{kek02} \\
${\alpha_\Lambda} (^5_\Lambda {\rm He})$ & $-0.25$ & $-0.60$
& $-0.80$  & $0.28$ ($0.24$)
 & $0.24 \pm 0.22$\cite{Ajim00} \\
\hline
$\Gamma (^{11}_\Lambda {\rm B})$ & $0.62$ & $0.36$
& $0.87$ & $0.88$ ($0.88$)
 &  $0.95  \pm 0.14$\cite{No95} \\
$n/p (^{11}_\Lambda {\rm B})$ & $0.10$ & $0.43$
 & $0.84$  & $0.92$ ($0.92$)
 & $1.04^{+0.59}_{-0.48}$\cite{Szy91} \\
${\cal A} (^{11}_\Lambda {\rm B})$ & $-0.09$ & $-0.22$
& $-0.22$  & $0.09$ ($0.08$)
 & $-0.20 \pm 0.10$\cite{Aj92} \\
\hline
$\Gamma (^{12}_\Lambda {\rm C})$ & $0.74$ & $0.41$
& $0.95$  & $0.93$ ($0.93$)
 & $1.14\pm 0.2$\cite{Szy91}, \, $0.89 \pm 0.15$\cite{No95}, \,\,
$0.83\pm 0.11$\cite{Bhang} \\
$n/p (^{12}_\Lambda {\rm C})$ & $0.08$ & $0.35$
& $0.67$  & $0.77$ ($0.77$)
 & $0.87 \pm 0.23$\cite{Ha02b} \\
${\cal A} (^{12}_\Lambda {\rm C})$ & $-0.03$ & $-0.06$
& $-0.05$ & $0.03$ ($0.02$)
 & $-0.01 \pm 0.10$\cite{Aj92} \\
\hline
${\hat \chi^2}$ & & & $0.93$ & $1.54$ ($1.15$) & \\
\end{tabular}
\end{table}


\begin{references}
\bibitem{ff}
While in principle a higher-order chiral loop calculation
would generate the effects of form factors term by term in the chiral
expansion, we will take the approach
that in the case of chiral SU(3) it is more effective to use a
{\sl phenomenological} prescription which
injects the ``right'' physics if the goal is to cleanly extract (or
codify) some phenomena rather than try to explain in detail
their origin\cite{ldr}.
The danger is, of course, that the {\sl phenomenological}
form factor might give
the correct physics on-shell or in the energy/momentum region wherein it was
designed to apply, and yet not be very accurate elsewhere.
\bibitem{comment1}
This is true for the $L=0$ part of the $\Lambda N$
wave function.
\bibitem{comment2}
Note that we are assuming
that ${\vec p}_1 - {\vec p}_2$ is small enough to disregard higher powers of
the derivative operators $ \vec{p}_1 - \vec{p}_2 $.
\end{references}
\end{document}